\documentstyle[12pt]{article}

\font\twlgot =eufm10 scaled \magstep1 \font\egtgot =eufm8
\font\sevgot =eufm7 \font\twlmsb =msbm10 scaled \magstep1
\font\egtmsb =msbm8 \font\sevmsb =msbm7

\newfam\gotfam
\def\pgot{\fam\gotfam\twlgot}
\textfont\gotfam\twlgot \scriptfont\gotfam\egtgot
\scriptscriptfont\gotfam\sevgot
\def\got{\protect\pgot}
\newfam\msbfam
\textfont\msbfam\twlmsb \scriptfont\msbfam\egtmsb
\scriptscriptfont\msbfam\sevmsb
\def\Bbb{\protect\pBbb}
\def\pBbb{\relax\ifmmode\expandafter\Bb\else\typeout{You cann't use
Bbb in text mode}\fi}
\def\Bb #1{{\fam\msbfam\relax#1}}

\newcommand{\gd}{{\got d}}
\newcommand{\ccG}{{\got g}}
\newcommand{\gA}{{\got A}}

\def\thebibliography#1{\section*{References}\list
  {[\arabic{enumi}]}{\settowidth\labelwidth{#1}\leftmargin\labelwidth
    \advance\leftmargin\labelsep
    \usecounter{enumi}}
    \def\newblock{\hskip .11em plus .33em minus .07em}
    \sloppy\clubpenalty4000\widowpenalty4000
    \sfcode`\.=1000\relax}

\def\op#1{\mathop{\fam0 #1}\limits}

\newcommand{\beq}{\begin{equation}}
\newcommand{\eeq}{\end{equation}}
\newcommand{\ben}{\begin{eqnarray}}
\newcommand{\een}{\end{eqnarray}}
\newcommand{\be}{\begin{eqnarray*}}
\newcommand{\ee}{\end{eqnarray*}}
\newcommand{\bea}{\begin{eqalph}}
\newcommand{\eea}{\end{eqalph}}
\newcommand{\cA}{{\cal A}}

\newcommand{\cP}{{\cal P}}
\newcommand{\cD}{{\cal D}}

\newcommand{\cV}{{\cal V}}
\newcommand{\cC}{{\cal C}}
\newcommand{\cL}{{\cal L}}
\newcommand{\cE}{{\cal E}}

\newcommand{\cM}{{\cal M}}
\newcommand{\cS}{{\cal S}}
\newcommand{\cO}{{\cal O}}
\newcommand{\bL}{{\bf L}}

\newcommand{\rL}{{\rm L}}

\newcommand{\al}{\alpha}

\newcommand{\bt}{\beta}
\newcommand{\dl}{\delta}
\newcommand{\la}{\lambda}
\newcommand{\La}{\Lambda}
\newcommand{\f}{\phi}
\newcommand{\om}{\omega}

\newcommand{\m}{\mu}
\newcommand{\n}{\nu}
\newcommand{\e}{\epsilon}
\newcommand{\g}{\gamma}
\newcommand{\G}{\Gamma}
\newcommand{\th}{\theta}

\newcommand{\vt}{\vartheta}
\newcommand{\vf}{\varphi}
\newcommand{\up}{\upsilon}

\newcommand{\di}{{\rm dim\,}}

\newcommand{\si}{\sigma}
\newcommand{\Si}{\Sigma}
\newcommand{\w}{\wedge}
\newcommand{\wt}{\widetilde}

\newcommand{\ol}{\overline}
\newcommand{\dr}{\partial}
\newcommand{\ar}{\op\longrightarrow}
\newcommand{\ot}{\otimes}
\newcommand{\ap}{\approx}
\newcommand{\ve}{\varepsilon}

\newcommand{\rdr}{\stackrel{\leftarrow}{\dr}{}}

\newcounter{theorem}
\newcounter{remark}
\newcounter{proposition}
\newcounter{lemma}
\newcounter{corollary}
\newcounter{definition}

\def\theremark{\arabic{remark}}

\def\thedefinition{\arabic{definition}}

\newcommand{\mar}[1]{}

\begin{document}
\hbox{}

\begin{center}

{\large\bf ON THE BV QUANTIZATION OF GAUGE GRAVITATION THEORY}
\bigskip

{\sc D.BASHKIROV}

{\it Department of Theoretical Physics, Moscow State University,
117234, Moscow, Russia \\ bashkir@phys.msu.ru}

\bigskip

{\sc G.SARDANASHVILY}

{\it Department of Theoretical Physics, Moscow State University,
117234, Moscow, Russia \\ sard@grav.phys.msu.su}

\end{center}

\bigskip

\noindent {\small  Quantization of gravitation theory as gauge
theory of general covariant transformations in the framework of
Batalin--Vilkoviski (BV) formalism is considered. Its gauge-fixed
Lagrangian is constructed. }

\section{Introduction}

Classical gauge gravitation theory has been developed in different
variants since 60th (see \cite{iva,heh,tmf} for a survey). We
describe gravitation theory as a standard gauge theory on fiber
bundles where parameters of gauge transformations are vector
fields on a base manifold $X$. The well-known BV quantization
technique \cite{bat,gom} can be applied to this theory. Since this
quantization technique fails to provide a functional measure, we
restrict ourselves to constructing a gauge-fixed Lagrangian.

Recall that an $r$-order Lagrangian of a Lagrangian system on a
smooth fiber bundle $Y\to X$  is defined as a density
\mar{0512}\beq
L=\cL\om:J^rY\to \op\w^nT^*X, \qquad \om=dx^1\w\cdots\w dx^n,
\qquad n=\di X, \label{0512}
\eeq
on the $r$-order jet manifold $J^rY$ of sections of $Y\to X$
\cite{book00,tenlect}. Given bundle coordinates $(x^\la,y^i)$ on
$Y$, the jet manifold $J^rY$ is endowed with the adapted
coordinates $(x^\la,y^i,y^i_\La)$, where $\La=(\la_k...\la_1)$,
$k=1,\ldots,r$, is a symmetric multi-index. We use the notation
$\la+\La=(\la\la_k...\la_1)$ and
\mar{5.177}\beq
d_\la = \dr_\la + \op\sum_{0\leq|\La|} y^i_{\la+\La}\dr_i^\La,
\qquad d_\La=d_{\la_r}\circ\cdots\circ d_{\la_1}. \label{5.177}
\eeq
A Lagrangian system on a fiber bundle is said to be a gauge system
if its Lagrangian $L$ admits a gauge symmetry depending on
parameter functions and their derivatives. In order to describe a
gauge system, let us consider Lagrangian formalism on the bundle
product
\mar{0681}\beq
E=Y\op \times_X W, \label{0681}
\eeq
where $W\to X$ is a vector bundle whose sections are gauge
parameter functions \cite{noether}. Let $W\to X$ be coordinated by
$(x^\la,\xi^r)$. Then a gauge symmetry is represented by a
differential operator
\mar{0509}\beq
\up= \op\sum_{0\leq|\La|\leq
m}\up^{i,\La}_r(x^\la,y^i_\Si)\xi^r_\La \dr_i \label{0509}
\eeq
on $E$ (\ref{0681}) which is linear on $W$ and takes its values in
the vertical tangent bundle $VY$ of $Y\to X$. By means of a
replacement of even gauge parameters $\xi^r$ and their jets
$\xi^r_\La$ with the odd ghosts $c^r$ and their jets $c^r_\La$,
the operator (\ref{0509}) defines a graded derivation
\mar{0680}\beq
\up= \op\sum_{0\leq|\La|\leq m}\up^{i,\La}_r(x^\la,y^i_\Si)c^r_\La
\dr_i \label{0680}
\eeq
of the algebra of the original even fields and odd ghosts. Its
extension
\mar{0684}\beq
\up= \op\sum_{0\leq|\La|\leq m}\up^{i,\La}_r c^r_\La \dr_i +
u^r\dr_r \label{0684}
\eeq
to ghosts is called the BRST transformation if it is nilpotent.
Such an extension exists if the original gauge transformations
form an algebra \cite{algebr}.

As was mentioned above, gauge transformations in gauge gravitation
theory are parameterized by vector fields on $X$. In this case,
the vector bundle $W\to X$ possesses the composite fibration $W\to
TX\to X$, where $TX$ is the tangent bundle of $X$. Here, we are
concerned with the following three gauge gravitation theories: (i)
the metric-affine gravitation theory on natural bundles describing
gravity in the absence of matter fields, (ii) gauge theory on
principal bundles with non-vertical gauge transformations which
describes gauge potentials of internal symmetries in the presence
of a gravitational field, (iii) gauge gravitation theory in the
presence of Dirac fermion fields. The corresponding gauge
transformations (\ref{gr3}), (\ref{i7}), (\ref{i15}), (\ref{i16}),
gauge symmetries (\ref{i21}), (\ref{gr11}), (\ref{i23}),
(\ref{i24}), and BRST symmetries (\ref{i30}), (\ref{i31}),
(\ref{i32}) are obtained. However, we apply the BV quantization
procedure only to the first of these gauge models describing pure
gravity. We construct its extended Lagrangian (\ref{i50}),
depending on ghost and antifields, and then the gauge-fixed
Lagrangian (\ref{i55}), where antifields are replaced with
gauge-fixing terms.

\section{Gauge gravitation theories}

Classical theory of gravity in the absence of matter fields can be
formulated as a gauge theory on natural bundles over an oriented
four-dimensional manifold $X$. The group of automorphisms of a
natural bundle contains the subgroup of general covariant
transformations defined as the canonical lift of diffeomorphisms
of $X$. They are gauge transformations of metric-affine
gravitation theory on natural bundles whose variables are linear
connections and pseudo-Riemannian metrics on $X$.

Linear connections on $X$ (henceforth world connection) are
connections on the principal bundle $LX$ of linear frames in the
tangent bundle $TX$ of $X$ with the structure group $GL_4=GL^+(4,
\Bbb R)$. These connections are represented by sections of the
quotient bundle
\mar{gr14}\beq
C_K=J^1LX/GL_4, \label{gr14}
\eeq
where $J^1LX$ is the first order jet manifold of sections of
$LX\to X$ \cite{book00,tenlect}. This fiber bundle is provided
with bundle coordinates $(x^\la,k_\la{}^\nu{}_\al)$ such that, for
any section $K$ of $C_K\to X$, its coordinates
$k_\la{}^\nu{}_\al\circ K=K_\la{}^\nu{}_\al$ are coefficient of
the linear connection
\be
K=dx^\la\ot (\dr_\la + K_\la{}^\m{}_\nu \dot x^\nu\dot\dr_\mu)
\ee
on $TX$ with respect to the holonomic bundle coordinates
$(x^\la,\dot x^\la)$. Note that the first order jet manifold
$J^1C_K$ of the fiber bundle (\ref{gr14}) possesses the canonical
decomposition taking the coordinate form
\mar{i52}\ben
&& k_{\la\m}{}^\al{}_\bt=\frac12(R_{\la\m}{}^\al{}_\bt +
S_{\la\m}{}^\al{}_\bt),\nonumber \\
&& R_{\la\m}{}^\al{}_\bt = k_{\la\m}{}^\al{}_\bt -
k_{\m\la}{}^\al{}_\bt + k_\m{}^\al{}_\ve k_\la{}^\ve{}_\bt
-k_\la{}^\al{}_\ve k_\m{}^\ve{}_\bt, \label{i52a}\\
&& S_{\la\m}{}^\al{}_\bt = k_{\la\m}{}^\al{}_\bt +
k_{\m\la}{}^\al{}_\bt - k_\m{}^\al{}_\ve k_\la{}^\ve{}_\bt
+k_\la{}^\al{}_\ve k_\m{}^\ve{}_\bt. \label{i52b}
\een
If $K$ is a section of $C_K\to X$, then $R\circ K$ is the
curvature of the world connection $K$.

In order to describe gravity, we assume that the linear frame
bundle $LX$ admits a Lorentz structure, i.e., a principal
subbundle $L^hX\subset LX$ whose structure group is the proper
Lorentz group $\rL=SO^0(1,3)$. By virtue of the well-known theorem
\cite{ste}, there is one-to-one correspondence between these
subbundles and the global sections $h$ of the quotient bundle
\mar{i0}\beq
\Si_T=LX/\rL. \label{i0}
\eeq
Namely, $L^hX=h^*LX$ is the bull-back of $LX\to\Si_T$ by $h$. We
agree to call $\Si_T$ the tetrad bundle. This is the twofold of
the metric bundle
\mar{i1}\beq
\Si=LX/SO(1,3), \label{i1}
\eeq
whose sections are pseudo-Riemannian (henceforth world) metrics on
$X$. Being an open subbundle of the tensor bundle $\op\vee^2 TX$,
the bundle $\Si$ (\ref{i1}) is provided with bundle coordinates
$(x^\la,\si^{\m\nu})$.

The configuration space of gauge gravitation theory in the absence
of matter fields is the bundle product $\Si\times C_K$. This is a
natural bundle admitting  the canonical lift
\mar{gr3}\ben
&& u_{K\Si}=u^\m\dr_\m +(\si^{\nu\bt}\dr_\nu u^\al
+\si^{\al\nu}\dr_\nu u^\bt)\frac{\dr}{\dr \si^{\al\bt}} +
\label{gr3}\\
&& \qquad (\dr_\nu u^\al k_\m{}^\nu{}_\bt -\dr_\bt u^\nu
k_\m{}^\al{}_\nu -\dr_\mu u^\nu k_\nu{}^\al{}_\bt
+\dr_{\m\bt}u^\al)\frac{\dr}{\dr k_\mu{}^\al{}_\bt} \nonumber
\een
of vector fields $u=u^\la\dr_\la$ on $X$ \cite{book00}. This lift
is a Lie algebra homomorphism, and the vector fields (\ref{gr3})
are infinitesimal generators of local one-dimensional groups of
general covariant transformations, whose gauge parameters are
vector fields on $X$.

Dealing with non-vertical automorphisms of non-natural bundles,
one also comes to gravitation theory. For instance, let us
consider the gauge theory  on a principal bundle $P\to X$ with a
structure Lie group $G$. In a general setting, the gauge-natural
prolongations of $P$ and the associated natural-gauge bundles are
called into play \cite{fat,fat04}. Principal connections on $P$
are represented by sections of the quotient
\mar{0654}\beq
C=J^1P/G\to X,\label{0654}
\eeq
called the bundle of principal connections \cite{book00}. This is
an affine bundle coordinated by $(x^\la, a^r_\la)$ such that,
given a section $A$ of $C\to X$, its components
$A^r_\la=a^r_\la\circ A$ are coefficients of the familiar local
connection form (i.e., gauge potentials). Infinitesimal generators
of local one-parameter groups of automorphisms of a principal
bundle $P$ are $G$-invariant projectable vector fields on $P\to
X$. They are represented by sections of the vector bundle
$T_GP=TP/G\to X$. This fiber bundle is endowed with the
coordinates $(x^\la,\tau^\la=\dot x^\la,\xi^r)$ with respect to
the fiber bases $\{\dr_\la, e_r\}$ for $T_GP$, where $\{e_r\}$ is
the basis for the right Lie algebra $\ccG$ of $G$ such that
$[e_p,e_q]=c^r_{pq}e_r.$ The bracket of sections of $T_GP\to X$
reads
\mar{0651}\ben
&& u_P=u^\la\dr_\la +u^r e_r, \qquad v_P=v^\la\dr_\la +v^r e_r,
\nonumber\\
 && [u_P,v_P]=(u^\m\dr_\m v^\la -v^\m\dr_\m u^\la)\dr_\la
+(u^\la\dr_\la v^r - v^\la\dr_\la u^r +c^r_{pq}u^pv^q)e_r.
\label{0651}
\een
Any section $u_P$ of the vector bundle $T_GP\to X$ yields the
vector field
\mar{0653}\beq
u_C=u^\la\dr_\la +(c^r_{pq}a^p_\la u^q +\dr_\la u^r -a^r_\m\dr_\la
u^\m)\dr^\la_r \label{0653}
\eeq
on the bundle of principal connections $C$ (\ref{0654}). For
instance, this vector field is a gauge symmetry of the global
Chern--Simons Lagrangian in gauge theory on a principal bundle
with a structure semi-simple Lie group $G$ over a
three-dimensional base $X$ \cite{chern}. In order to obtain a
gauge symmetry of the Yang--Mills Lagrangian, one have to extend
the vector field (\ref{0653}) to the vector field
\mar{i7}\beq
u_{C\Si}=u^\la\dr_\la +(c^r_{pq}a^p_\la u^q +\dr_\la u^r
-a^r_\m\dr_\la u^\m)\dr^\la_r  + (\si^{\nu\bt}\dr_\nu u^\al
+\si^{\al\nu} \dr_\nu u^\bt)\frac{\dr}{\dr \si^{\al\bt}}
\label{i7}
\eeq
on the bundle product $C\times \Si$ of $C$ and the metric bundle
$\Si$ (\ref{i1}).

The physical underlying reasons for the existence of a Lorentz
structure and, consequently, a world metric are Dirac fermion
fields. They are described by sections of Dirac spinor bundles,
which fail to be natural. In order to develop gauge theory of
gravity in the presence of Dirac fermion fields, one has to define
general covariant transformations of these fields.

Recall that Dirac spinors are described in terms of Clifford
algebras as follows \cite{cra,obu}. Let $M$ be the Minkowski space
equipped with the Minkowski metric $\eta$ of signature $(+---)$,
and let $\{t^a\}$ be a fixed basis for $M$. The complex Clifford
algebra $\Bbb C_{1,3}$ is defined as the complexified quotient of
the tensor algebra $\ot M$ of $M$ by the two-sided ideal generated
by elements
\be
t\otimes t'+t'\otimes t-2\eta(t,t')\in \ot M,\qquad t,t'\in M.
\ee
The Dirac spinor space $V_s$ is introduced as a minimal left ideal
of $\Bbb C_{1,3}$  on which this algebra acts by left
multiplications. There is the representation
\mar{w01}\beq
\g: M\otimes V \to V, \qquad \g(t^a)=\g^a, \label{w01}\\
\eeq
of elements of the Minkowski space $M\subset{\Bbb C}_{1,3}$ by the
Dirac $\g$-matrices on $V_s$. Different ideals $V_s$ lead to
equivalent representations (\ref{w01}). The complex Clifford
algebra $\Bbb C_{1,3}$ is isomorphic to the real Clifford algebra
$\Bbb R_{2,3}$, whose generating space is $\Bbb R^5$ endowed with
the metric of signature $(+---+)$. Its subalgebra generated by
elements of $M\subset \Bbb R^5$ is the real Clifford algebra $\Bbb
R_{1,3}$. The Clifford group $G_{1,3}$ consists of the invertible
elements $l_s$ of $\Bbb R_{1,3}$ such that the inner automorphisms
defined by these elements preserve the Minkowski space $M\subset
\Bbb R_{1,3}$. Since this action of $G_{1,3}$ on $M$ is not
effective, one consider its spin subgroup $\rL_s= SL(2,\Bbb C)$,
acting on the spinor space $V_s$ by means of the generators
\mar{b3213}\beq
L_{ab}=\frac{1}{4}[\g_a,\g_b]. \label{b3213}
\eeq
This action preserves the representation (\ref{w01}) and the
spinor metric
\mar{b3210}\beq
a(v,v')=\frac12(v^+\g^0v' +{v'}^+\g^0v), \qquad v,v'\in V_s.
\label{b3210}
\eeq

The spin group $\rL_s$ is the universal twofold  of the proper
Lorentz group L. Given the universal twofold $\wt{GL}_4$ of the
group $GL_4$, we have the commutative diagram
\mar{b3243}\beq
\begin{array}{ccc}
 \wt{GL}_4 & \longrightarrow &  GL_4 \\
 \put(0,-10){\vector(0,1){20}} &
& \put(0,-10){\vector(0,1){20}}  \\
\rL_s & \ar &  \rL
\end{array} \label{b3243}
\eeq
Note that the group $\wt{GL}_4$ admits spinor representations, but
they are infinite-dimensional \cite{heh,nee}. Let us consider the
$\wt{GL}$-principal bundle $\wt{LX}\to X$ which is the twofold of
the linear frame bundle $LX$ \cite{dabr,law,swit}. We agree to
call it the spin frame bundle. One says that $\wt{LX}$ admits a
Dirac spin structure if there exists its $\rL_s$-principal
subbundle $P^h$. Then we have the commutative diagram
\mar{b3222}\beq
\begin{array}{ccc}
 \wt{LX} & \ar^\zeta & \wt{LX} \\
 \put(0,-10){\vector(0,1){20}} &
& \put(0,-10){\vector(0,1){20}}  \\
P^h & \ar & L^hX
\end{array} \label{b3222}
\eeq
where $L^hX$ is some Lorentz structure \cite{fulp,sard98a}. As a
consequence, $\wt{LX}/\rL_s=\Si_T$ (\ref{i0}), and there is
one-to-one correspondence $P^h=h^*\wt{LX}$ between the Dirac spin
structures $P^h\subset\wt{LX}$ and the global sections $h$ of the
tetrad bundle $\Si_T\to X$. We agree to call these sections the
tetrad fields because of the following.

With respect to the holonomic frames $\{\dr_\m\}$ in the tangent
bundle $TX$, every element $\{H_a\}$ of the linear frame bundle
$LX$ takes the form $H_a=H^\m_a\dr_\m$. The matrix elements
$H^\m_a$ constitute the bundle coordinates on $LX\to X$ such that
the transition functions and the canonical right action of $GL_4$
on $LX$ take the form
\be
H'^\m_a=\frac{\dr x'^\m}{\dr x^\la}H^\la_a, \qquad GL_4\ni g:
H^\m_a\mapsto H^\m_bg^b{}_a.
\ee
Let collections of local sections $\wt\Psi=\{z^s_\e\}$ and
$\Psi=\{z_\e=\zeta\circ z^s_\e\}$ yield associated atlases of the
principal bundles $\wt{LX}\to \Si_T$ and $LX\to\Si_T$,
respectively. Then the tetrad bundle $\Si_T\to X$ can be provided
with bundle coordinates $(x^\la, \si_a^\m= H^\m_a\circ \zeta_e)$.
Components of a tetrad field $h$ with respect to these coordinates
are tetrad functions $h^\m_a=\si^a_\m\circ h$. The familiar
relation $g^{\m\nu}=h^\m_ah^\nu_b\eta_{ab}$ between tetrad
functions and metric functions of the world metric
$g:X\ar^h\Si_T\to\Si$ holds.

There is the well-known topological obstruction to the existence
of a Dirac spin structure. Let us restrict our consideration to
non-compact manifolds $X$. Then a Dirac spinor structure exists
iff $X$ is parallelizable \cite{ger} (e.g., $X=\Bbb R\times Z$
because any orientable three-dimensional manifold $Z$ is
parallelizable \cite{sti}). In this case, the spin frame bundle is
unique and all Dirac spin structures are equivalent
\cite{ger,avis}, i.e., all $\rL_s$-principal subbundles $P^h$ of
$\wt{LX}$ are pairwise isomorphic by means of automorphisms of
$\wt{LX}$ and isomorphic to the trivial $\rL_s$-principal bundle
$P^0\to X$.

Since $\wt{LX}\to \Si_T$ is an $\rL_s$-principal bundle, one can
consider the associated spinor bundle
\mar{yy1}\beq
S=(\wt{LX}\times V_s)/\rL_s\to\Si_T \label{yy1}
\eeq
over $\Si_T$ \cite{book00,sard98a}. We agree to call it the
universal spinor bundle because, given a tetrad field $h$, the
pull-back $S^h=h^*S\to X$ of $S$ onto $X$ is a spinor bundle on
$X$ which is associated to the $\rL_s$-principal bundle $P^h$.
Sections of $S^h$ describe Dirac fermion fields in the presence of
a tetrad gravitational field $h$. Given an atlas $\wt\Psi$ of
$\wt{LX}$ and the corresponding bundle coordinates $(x^\la,
\si^\m_a)$ on $\Si_T$, the universal spinor bundle $S$ (\ref{yy1})
is endowed with bundle coordinates $(x^\la, \si^\m_a,y^A)$, where
$y^A$ are coordinates on the spinor space $V_s$.

The universal spinor bundle $S\to\Si_T$ is a subbundle of the
bundle of Clifford algebras which is generated by the bundle of
Minkowski spaces
\mar{mos266}\beq
E_M=(LX\times M)/\rL\to\Si_T \label{mos266}
\eeq
associated with the L-principal bundle $LX\to\Si_T$.  Then due to
the representation $\g$ (\ref{w01}), there exists the
representation
\mar{L7}\beq
\g_\Si: T^*X\op\ot_{\Si_T} S\ar (\wt{LX}\times (M\ot V))/\rL_s \to
(\wt{LX}\times\g(M\ot V))/\rL_s=S, \label{L7}
\eeq
defined by the coordinate expression $\g_\Si (dx^\la)
=\si^\la_a\g^a$.

Given a world connection $K$ on $X$, the universal spinor bundle
$S\to\Si_T$ is provided with the connection
\mar{b3266}\ben
&& A_\Si = dx^\la\ot(\dr_\la - \frac14
(\eta^{kb}\si^a_\m-\eta^{ka}\si^b_\m)
 \si^\nu_k K_\la{}^\m{}_\nu L_{ab}{}^A{}_By^B\dr_A) +
 \label{b3266}\\
&& \qquad d\si^\m_k\ot(\dr^k_\m + \frac14 (\eta^{kb}\si^a_\m
-\eta^{ka}\si^b_\m) L_{ab}{}^A{}_By^B\dr_A). \nonumber
\een
Given a tetrad field $h$, its restriction to $S^h$ is the familiar
spin connection
\mar{b3212}\beq
K_h=dx^\la\ot[\dr_\la +\frac14
(\eta^{kb}h^a_\m-\eta^{ka}h^b_\m)(\dr_\la h^\m_k - h^\nu_k
K_\la{}^\m{}_\nu)L_{ab}{}^A{}_B y^B\dr_A], \label{b3212}
\eeq
defined by a world connection $K$ \cite{pon,sard97b}. The
connection (\ref{b3266}) yields the first order differential
operator
\mar{7.10'}\ben
&& D: :J^1S\to T^*X\op\ot_{\Si_T} S, \nonumber\\
&& D= dx^\la\ot[y^A_\la- \frac14(\eta^{kb}\si^a_\m
-\eta^{ka}\si^b_\m)(\si^\m_{\la k} -\si^\nu_k
K_\la{}^\m{}_\nu)L_{ab}{}^A{}_By^B]\dr_A, \label{7.10'}
\een
on the fiber bundle $S\to X$. Its restriction to $J^1S^h\subset
J^1S$ recovers the familiar covariant differential on the spinor
bundle $S^h\to X$ relative to the spin connection (\ref{b3212}).
Combining (\ref{L7}) and (\ref{7.10'}) gives the first order
differential operator
\mar{3261}\ben
&& \cD=\g_\Si\circ D:J^1S\to T^*X\op\ot_{\Si_T}
S\to S, \label{b3261}\\
&& \cD=\si^\la_a\g^{aB}{}_A[y^A_\la- \frac14(\eta^{kb}\si^a_\m
-\eta^{ka}\si^b_\m)(\si^\m_{\la k} -\si^\nu_k
K_\la{}^\m{}_\nu)L_{ab}{}^A{}_By^B], \nonumber
\een
on the fiber bundle $S\to X$. Its restriction to $J^1S^h\subset
J^1S$ is the familiar Dirac operator on the spinor bundle $S^h$ in
the presence of a background tetrad field $h$ and a world
connection $K$.

Thus, we come to the model of the metric-affine gravity and Dirac
fermion fields. Its configuration space is the bundle product
\mar{042}\beq
Y=C_K \op\times_{\Si_T} S, \label{042}
\eeq
coordinated by $(x^\m,\si^\m_a, k_\m{}^\al{}_\bt,y^A)$. Let
$J^1_{\Si_T}Y$ denote the first order jet manifold of the fiber
bundle $Y\to \Si_T$. This fiber bundle can be endowed with the
spin connection
\mar{b3263}\ben
&& A_Y= dx^\la\ot(\dr_\la - \frac14
(\eta^{kb}\si^a_\m-\eta^{ka}\si^b_\m)
 \si^\nu_k k_\la{}^\m{}_\nu L_{ab}{}^A{}_By^B\dr_A) + \label{b3263}\\
&& \qquad d\si^\m_k\ot(\dr^k_\m + \frac14 (\eta^{kb}\si^a_\m
-\eta^{ka}\si^b_\m) L_{ab}{}^A{}_By^B\dr_A).\nonumber
\een
With this connection, we obtain the first order differential
operator
\mar{7.100}\beq
D_Y=dx^\la\ot[y^A_\la- \frac14(\eta^{kb}\si^a_\m
-\eta^{ka}\si^b_\m)(\si^\m_{\la k} -\si^\nu_k
k_\la{}^\m{}_\nu)L_{ab}{}^A{}_By^B]\dr_A, \label{7.100}
\eeq
and the total Dirac operator
\mar{b3264}\beq
 \cD_Y=\si^\la_a\g^{aB}{}_A[y^A_\la-
\frac14(\eta^{kb}\si^a_\m -\eta^{ka}\si^b_\m)(\si^\m_{\la k}
-\si^\nu_k k_\la{}^\m{}_\nu)L_{ab}{}^A{}_B y^B], \label{b3264}
\eeq
on  the fiber bundle $Y\to X$. Given a world connection $K:X\to
C_K$, the restrictions of the spin connection $A_Y$ (\ref{b3263}),
the operators $D_Y$ (\ref{7.100}) and $\cD_Y$ (\ref{b3264}) to
$K^*Y$ are exactly the spin connection (\ref{b3266}), the
operators (\ref{7.10'}) and (\ref{b3261}), respectively. The total
Lagrangian of the metric-affine gravity and fermion fields is the
sum of a metric-affine Lagrangian $L_{\rm MA}$ depending on the
variables $\si^{\m\n}=\si^\m_a\si^\nu_b\eta^{ab}$,
$k_\m{}^\al{}_\bt$ and the Dirac Lagrangian constructed from the
Dirac operator (\ref{b3264}) and the spinor metric (\ref{b3210})
\cite{book00,sard98a}.

Turn now to gauge transformations in this gauge theory. The spin
frame bundle $\wt{LX}\to X$ as like as the linear frame one $LX\to
X$ is natural, and it admits the canonical lift of any vector
field on $X$. Hence, the universal spinor bundle $S\to X$ is also
natural, and the canonical lift $u_S$ onto $S$ of vector fields on
$X$ exists. The goal is to find its coordinate expression.
Difficulties arise because the tetrad coordinates $\si^\m_a$ on
$\Si_T$ depend on the choice of an atlas $\Psi$ of the bundle
$LX$. Therefore, non-canonical components appear in the coordinate
expression of $u_S$ \cite{sard98a,sard97b}.

Since the fiber bundles $LX\to X$, $GL_4\to GL_4/\rL$ and,
accordingly, the fiber bundles $\wt{LX}\to X$, $\wt{GL}_4\to\rL_s$
are trivial, the fiber bundles $\wt{LX}\to\Si_T$ and $S\to\Si_T$
are also trivial. Let $\Psi$ be an atlas of $\wt{LX}\to \Si_T$
with transition functions constant on fibers of $\Si_T$. This
atlas provides a trivialization  $\wt{LX}\cong P^0\op\times \Si_T$
and the corresponding trivialization
\mar{i10}\beq
S\cong S^0\op\times \Si_T \label{i10}
\eeq
of the universal spinor bundle. With respect to such an atlas, one
can define the lift
\mar{i11}\ben
&& u_S=u^\la\dr_\la + \dr_\nu u^\al\si^\nu_c \dr^c_\al + \label{i11}\\
&& \qquad \frac14 (\eta^{kb}\si^a_\m-\eta^{ka}\si^b_\m)
 (\dr_\nu u^\m \si^\nu_k- u^\nu\si^\m_{\nu k})(-L_{ab}{}^d{}_c \si^\al_d \dr^c_\al +
 L_{ab}{}^A{}_By^B\dr_A),  \nonumber\\
&& \qquad L_{ab}{}^d{}_c=\eta_{bc}\dl^d_a-\eta_{ac}\dl^d_b,
\nonumber
\een
onto $S$ of vector fields $u=u^\la\dr_\la$ on $X$. This lift
however is not a vector field because it depends on the jet
coordinates $\si^\m_{\nu k}$. It is a generalized vector field
(see the next Section).  Its projection onto $S^h$ in accordance
with the splitting (\ref{i10}) is the well-known Lie derivative
\mar{i12}\beq
 u_h=u^\la\dr_\la + \frac14 (\eta^{kb}h^a_\m-\eta^{ka}h^b_\m)
 (\dr_\nu u^\m h^\nu_k- u^\nu\dr_\nu h^\m_k)L_{ab}{}^A{}_By^B\dr_A
\label{i12}
\eeq
of spinor fields in the presence of a tetrad field $h$
\cite{kosm,fat98}. Note that the lift (\ref{i11}) fails to be a
Lie algebra homomorphism.

At the same time, one can think of the fiber bundle $S\to X$
(\ref{i10}) as being associated with the $\rL_s$-principal bundle
$P^0\to X$. Infinitesimal generators of local one-parameter groups
of automorphisms of $P^0$ are represented by sections of the
vector bundle $TP^0/\rL_s\to X$ which take the coordinate form
$u_P=u^\la\dr_\la +u^me_m$, where $\{e_m\}$ is the basis for the
Lie algebra of $\rL_s$. Any vector field $u_P$ yields the vector
field
\mar{i13}\beq
u_S=u^\la\dr_\la + (\dr_\nu u^\al\si^\nu_c -u^m
I_m{}^d{}_c\si^\al_d) \dr^c_\al + u^m I_m{}^A{}_B y^B\dr_A
\label{i13}
\eeq
on the associated bundle $S$. These vector fields can also be
regarded as infinitesimal gauge transformations of gauge theory of
gravity in the presence of Dirac fermion fields. The total
configuration space of this theory is the bundle product
\mar{i14}\beq
Y=S\op\times_{\Si_T} C_K \label{i14}
\eeq
coordinated by $(x^\la,\si^\m_a, k_\m{}^\al{}_\bt, y^A)$. The
generalized vector fields (\ref{i11}) and the vector fields
(\ref{i13}) are extended to the fiber bundle (\ref{i14}) as
\mar{i15,6}\ben
&& u_{SK}=u^\la\dr_\la + \dr_\nu u^\al\si^\nu_c \dr^c_\al + \label{i15}\\
&& \qquad \frac14 (\eta^{kb}\si^a_\m-\eta^{ka}\si^b_\m)
 (\dr_\nu u^\m \si^\nu_k- u^\nu\si^\m_{\nu k})(-L_{ab}{}^d{}_c \si^\al_d \dr^c_\al +
 L_{ab}{}^A{}_By^B\dr_A)+  \nonumber\\
&& \qquad [\dr_\nu u^\al k_\m{}^\nu{}_\bt -\dr_\bt u^\nu
k_\m{}^\al{}_\nu -\dr_\mu u^\nu k_\nu{}^\al{}_\bt
+\dr_{\m\bt}u^\al]\frac{\dr}{\dr k_\mu{}^\al{}_\bt}, \nonumber\\
&& u_{SK}=u^\la\dr_\la + (\dr_\nu u^\al\si^\nu_c -u^m
I_m{}^d{}_c\si^\al_d) \dr^c_\al + u^m I_m{}^A{}_B y^B\dr_A +
\label{i16}\\
&& \qquad [\dr_\nu u^\al k_\m{}^\nu{}_\bt -\dr_\bt u^\nu
k_\m{}^\al{}_\nu -\dr_\mu u^\nu k_\nu{}^\al{}_\bt
+\dr_{\m\bt}u^\al]\frac{\dr}{\dr k_\mu{}^\al{}_\bt}.\nonumber
\een

\section{Gauge symmetries}

In a general setting, gauge systems on fiber bundle $Y\to X$ are
described as follows \cite{noether}.

With the inverse system of jet manifolds
\mar{5.10}\beq
X\op\longleftarrow^\pi Y\op\longleftarrow^{\pi^1_0} J^1Y
\longleftarrow \cdots J^{r-1}Y \op\longleftarrow^{\pi^r_{r-1}}
J^rY\longleftarrow\cdots, \label{5.10}
\eeq
one has the direct system
\mar{5.7}\beq
\cO^*X\op\longrightarrow^{\pi^*} \cO^*Y
\op\longrightarrow^{\pi^1_0{}^*} \cO_1^*Y \ar\cdots \cO^*_{r-1}Y
\op\longrightarrow^{\pi^r_{r-1}{}^*}
 \cO_r^*Y \longrightarrow\cdots \label{5.7}
\eeq
of graded differential algebras (henceforth GDAs) $\cO_r^*Y$ of
exterior forms on jet manifolds $J^rY$ with respect to the
pull-back monomorphisms $\pi^r_{r-1}{}^*$. Its direct limit
 $\cO_\infty^*Y$ is a GDA
consisting of all exterior forms on finite order jet manifolds
modulo the pull-back identification.

The projective limit $(J^\infty Y, \pi^\infty_r:J^\infty Y\to
J^rY)$ of the inverse system (\ref{5.10}) is a Fr\'echet manifold.
A bundle atlas $\{(U_Y;x^\la,y^i)\}$ of $Y\to X$ yields the
coordinate atlas
\mar{jet1}\beq
\{((\pi^\infty_0)^{-1}(U_Y); x^\la, y^i_\La)\}, \qquad
{y'}^i_{\la+\La}=\frac{\dr x^\m}{\dr x'^\la}d_\m y'^i_\La, \qquad
0\leq|\La|, \label{jet1}
\eeq
of $J^\infty Y$, where $d_\m$ are the total derivatives
(\ref{5.177}).  Then $\cO^*_\infty Y$ can be written in a
coordinate form where the horizontal one-forms $\{dx^\la\}$ and
the contact one-forms $\{\th^i_\La=dy^i_\La
-y^i_{\la+\La}dx^\la\}$ are local generating elements of the
$\cO^0_\infty Y$-algebra $\cO^*_\infty Y$. There is the canonical
decomposition $\cO^*_\infty Y=\oplus\cO^{k,m}_\infty Y$ of
$\cO^*_\infty Y$ into $\cO^0_\infty Y$-modules $\cO^{k,m}_\infty
Y$ of $k$-contact and $m$-horizontal forms together with the
corresponding projectors $h_k:\cO^*_\infty Y\to \cO^{k,*}_\infty
Y$. Accordingly, the exterior differential on $\cO_\infty^* Y$ is
split into the sum $d=d_H+d_V$ of the nilpotent total and vertical
differentials
\be
d_H(\f)= dx^\la\w d_\la(\f), \qquad d_V(\f)=\th^i_\La \w
\dr^\La_i\f, \qquad \f\in\cO^*_\infty Y.
\ee

Any finite order Lagrangian (\ref{0512}) is an element of
$\cO^{0,n}_\infty Y$, while
\mar{0513}\beq
\dl L=\cE_i\th^i\w\om=\op\sum_{0\leq|\La|}
(-1)^{|\La|}d_\La(\dr^\La_i \cL)\th^i\w\om\in \cO^{1,n}_\infty Y
\label{0513}
\eeq
is its Euler--Lagrange operator taking values into the vector
bundle
\mar{0548}\beq
V^*Y\op\ot_Y\op\w^n T^*X, \label{0548}
\eeq
where $V^*Y$ denotes the vertical cotangent bundle of $Y\to X$.
Given a Lagrangian $L$ and its Euler-Lagrange operator $\dl L$
(\ref{0513}), we further abbreviate $A\ap 0$ with an equality
which holds on-shell. This means that $A$ is an element of a
module over the ideal $I_L$ of the ring $\cO^0_\infty Y$ which is
locally generated by the variational derivatives $\cE_i$ and their
total derivations $d_\La\cE_i$. We say that $I_L$ is a
differential ideal because, if a local function $f$ belongs to
$I_L$, then every total derivative $d_\La f$ does as well.

A Lagrangian system on a fiber bundle $Y\to X$ is said to be a
gauge theory if its Lagrangian $L$ admits a family of variational
symmetries parameterized by elements of a vector bundle $W\to X$
and its jet manifolds as follows.

Let $\gd\cO^0_\infty Y$ be the $\cO^0_\infty Y$-module of
derivations of the $\Bbb R$-ring $\cO^0_\infty Y$. Any $\vt\in
\gd\cO^0_\infty Y$ yields the graded derivation (the interior
product) $\vt\rfloor\f$ of the GDA $\cO^*_\infty Y$ given by the
relations
\be
\vt\rfloor df=\vt(f), \qquad  \vt\rfloor(\f\w\si)=(\vt\rfloor
\f)\w\si +(-1)^{|\f|}\f\w(\vt\rfloor\si), \quad f\in \cO^0_\infty
Y, \quad \f,\si\in \cO^*_\infty Y,
\ee
and its derivation (the Lie derivative)
\mar{0515}\beq
\bL_\vt\f=\vt\rfloor d\f+ d(\vt\rfloor\f), \qquad
\bL_\vt(\f\w\f')=\bL_\vt(\f)\w\f' +\f\w\bL_\vt(\f'), \qquad
\f,\f'\in \cO^*_\infty Y. \label{0515}
\eeq
Relative to an atlas (\ref{jet1}), a derivation
$\vt\in\gd\cO^0_\infty$ reads
\mar{g3}\beq
\vt=\vt^\la \dr_\la + \vt^i\dr_i + \op\sum_{|\La|>0}\vt^i_\La
\dr^\La_i, \label{g3}
\eeq
where the tuple of derivations $\{\dr_\la,\dr^\La_i\}$ is defined
as the dual of that of the exterior forms $\{dx^\la, dy^i_\La\}$
with respect to the interior product $\rfloor$ \cite{cmp}.

A derivation $\vt$ is called contact if the Lie derivative
$\bL_\vt$ (\ref{0515}) preserves the contact ideal of the GDA
$\cO^*_\infty Y$ generated by contact forms. A derivation $\up$
(\ref{g3}) is contact iff
\mar{g4}\beq
\vt^i_\La=d_\La(\vt^i-y^i_\m\vt^\m)+y^i_{\m+\La}\vt^\m, \qquad
0<|\La|. \label{g4}
\eeq
Any contact derivation admits the horizontal splitting
\mar{g5}\beq
\vt=\vt_H +\vt_V=\vt^\la d_\la + (\up^i\dr_i + \op\sum_{0<|\La|}
d_\La \up^i\dr_i^\La), \qquad \up^i= \vt^i-y^i_\m\vt^\m.
\label{g5}
\eeq
Its vertical part $\vt_V$  is completely determined by the first
summand
\mar{0641}\beq
\up=\up^i(x^\la,y^i_\La)\dr_i, \qquad 0\leq |\La|\leq k.
\label{0641}
\eeq
This is a section of the pull-back $VY\op\times_Y J^kY\to J^kY$,
i.e., a $k$-order $VY$-valued differential operator on $Y$. One
calls $\up$ (\ref{0641}) a generalized vector field on $Y$.

One can show that the Lie derivative of a Lagrangian $L$
(\ref{0512}) along a contact derivation $\vt$ (\ref{g5}) fulfills
the first variational formula
\mar{g8'}\beq
\bL_\vt L= \up\rfloor\dl L +d_H(h_0(\vt\rfloor\Xi_L)) +\cL d_V
(\vt_H\rfloor\om), \label{g8'}
\eeq
where $\Xi_L$ is a Lepagean equivalent of $L$ \cite{cmp}. A
contact derivation $\vt$ (\ref{g5}) is called  variational if the
Lie derivative (\ref{g8'}) is $d_H$-exact, i.e., $\bL_\vt
L=d_H\si$, $\si\in \cO^{0,n-1}_\infty$. A glance at the expression
(\ref{g8'}) shows that: (i) $\vt$ (\ref{g5}) is variational only
if it is projected onto $X$; (ii) $\vt$ is variational iff its
vertical part $\vt_V$ is well; (iii) it is variational iff
$\up\rfloor\dl L$ is $d_H$-exact. Therefore, we can restrict our
consideration to vertical contact derivations $\vt=\vt_V$. A
generalized vector field $\up$ (\ref{0641}) is called a
variational symmetry of a Lagrangian $L$ if it generates a
variational contact derivation.

In order to define a gauge symmetry \cite{noether}, let us
consider the bundle product $E$ (\ref{0681}) coordinated by
$(x^\la,y^i,\xi^r)$. Given a Lagrangian $L$ on $Y$, let us
consider its pull-back, say again $L$, onto $E$. Let $\vt_E$ be a
contact derivation of the $\Bbb R$-ring $\cO^0_\infty E$, whose
restriction
\mar{0508}\beq
\vt=\vt_E|_{\cO^0_\infty Y}=
\op\sum_{0\leq|\La|}d_\La\up^i\dr_i^\La \label{0508}
\eeq
to $\cO^0_\infty Y\subset \cO^0_\infty E$ is linear in coordinates
$\xi^r_\Xi$. It is determined by a generalized vector field
$\up_E$ on $E$ whose projection
\be
\up:J^kE\ar^{\up_E} VE\to E\op\times_Y VY
\ee
is a linear $VY$-valued differential operator $\up$ (\ref{0509})
on $E$. Let $\vt_E$ be a variational symmetry of a Lagrangian $L$
on $E$, i.e.,
\mar{0552}\beq
\up_E\rfloor \dl L=\up\rfloor \dl L=d_H\si. \label{0552}
\eeq
Then one says that $\up$ (\ref{0509}) is a gauge symmetry of a
Lagrangian $L$.

In accordance with Noether's second theorem \cite{noether}, if a
Lagrangian $L$ (\ref{0512}) admits a gauge symmetry $\up$
(\ref{0509}),  its Euler--Lagrange operator (\ref{0513}) obeys the
Noether identity
\mar{0550}\ben
&& \Delta\circ \dl L=0, \label{0550}\\
&&  [\op\sum_{0\leq|\La|\leq m} \Delta^{i,\La}_r d_\La \cE_i]
\xi^r\om=[\op\sum_{0\leq|\La|\leq
m}(-1)^{|\La|}d_\La(\up^{i,\La}_r\cE_i)]\xi^r\om=0, \nonumber
\een
where
\mar{0511}\ben
&& \Delta= \xi^r\Delta_r\om= \op\sum_{0\leq|\La|\leq m}
\xi^r\Delta^{i,\La}_r(x^\la,y^j_\Si)\ol y_{\La i} \om=
\xi^r[\op\sum_{0\leq|\La|\leq
m}(-1)^{|\La|}d_\La(\up^{i,\La}_r\ol y_i)]\om, \label{0511}\\
&& \Delta^{i,\La}_r =\op\sum_{0\leq|\Si|\leq
m-|\La|}(-1)^{|\Si+\La|}C^{|\Si|}_{|\Si+\La|} d_\Si
\up^{i,\Si+\La}_r,\nonumber
\een
is the Noether operator on the fiber bundle (\ref{0548}),
coordinated by $(x^\la,y^i, \ol y_i)$, with values in the fiber
bundle
\mar{0630}\beq
E^*\op\ot_Y\op\w^n T^*X. \label{0630}
\eeq

For instance, if a gauge symmetry
\mar{0656}\beq
\up=(\up_r^i\xi^r +\up^{i,\m}_r\xi^r_\m
+\up_r^{i,\nu\m}\xi^r_{\nu\m})\dr_i \label{0656}
\eeq
is of second jet order in parameters,  the corresponding Noether
operator (\ref{0511})  reads
\mar{0657}\beq
\Delta^i_r=\up^i_r -d_\m \up^{i,\m}_r
+d_{\nu\m}\up_r^{i,\nu\m},\qquad \Delta^{i,\m}_r=- \up^{i,\m}_r
+2d_\nu\up_r^{i,\nu\m}, \qquad \Delta_r^{i,\nu\m}=\up_r^{i,\nu\m}.
\label{0657}
\eeq

\section{Gauge symmetries in gauge gravitation theories}

Turn now to gauge symmetries of gauge gravitation theories
mentioned in Section 2.

Let start with the metric-affine gravitation theory on the fiber
bundle $\Si\times C_K$ with the infinitesimal gauge
transformations (\ref{gr3}) parameterized by vector fields on $X$.
In this case, the configuration space (\ref{0681}) of the gauge
system is the bundle product
\mar{i20}\beq
E=\Si\op\times_X C_K\op\times_X TX, \label{i20}
\eeq
coordinated by $(x^\la,\si^{\m\nu}, k_\la{}^\nu{}_\al,\tau^\la)$,
and the gauge symmetry (\ref{0509}) is the generalized vector
field
\mar{i21}\ben
&& \up=(\tau_\nu^\al \si^{\nu\bt} + \tau_\nu^\bt\si^{\al\nu}
-\tau^\nu\si^{\al\bt}_\nu)\frac{\dr}{\dr \si^{\al\bt}} +
\label{i21}\\
&& \qquad (\tau_\nu^\al k_\m{}^\nu{}_\bt -\tau_\bt^\nu
k_\m{}^\al{}_\nu -\tau_\mu^\nu k_\nu{}^\al{}_\bt +\tau_{\m\bt}^\al
-\tau^\nu k_{\nu\m}{}^\al{}_\bt)\frac{\dr}{\dr k_\mu{}^\al{}_\bt}.
\nonumber
\een

Let us consider the gauge theory of principal connections in the
presence of a metric gravitational field on the fiber bundle
$C\times\Si$ with the infinitesimal gauge transformations
(\ref{i7}). The configuration space of the corresponding gauge
system is the bundle product
\mar{gr10}\beq
E=C\op\times_X \Si\op\times_X T_GP, \label{gr10}
\eeq
coordinated by $(x^\la,a^r_\la,\si^{\al\bt}, \tau^\la, \xi^r)$,
and its gauge symmetry is given by the generalized vector field
\mar{gr11}\beq
\up=(c^r_{pq}a^p_\la \xi^q + \xi^r_\la -\tau^\m_\la a^r_\m
-\tau^\m a_{\m\la}^r)\dr^\la_r + (\tau_\nu^\al\si^{\nu\bt} +
\tau_\nu^\bt\si^{\al\nu} -\tau^\la\si_\la^{\al\bt})\frac{\dr}{\dr
\si^{\al\bt}}. \label{gr11}
\eeq

The configuration space of the gauge theory of gravity in the
presence of Dirac fermion fields is the fiber bundle (\ref{i14}).
In the case of infinitesimal gauge transformations (\ref{i15}),
the corresponding gauge system is defined on the bundle product
\mar{i22}\beq
E=S\op\times_{\Si_T} C_K \op\times_{\Si_T}TX, \label{i22}
\eeq
coordinated by $(x^\la,\si^\m_a, k_\la{}^\nu{}_\al,
y^A,\tau^\la)$. Its gauge symmetry reads
\mar{i23}\ben
&& \up= \frac12(\eta^{kd}\eta_{ac}\si^a_\m\si^\al_d +\dl^\al_\m\dl^k_c)
(\tau_\nu^\m \si^\nu_k- \tau^\nu\si^\m_{\nu k}) \dr^c_\al + \label{i23}\\
&& \qquad [\frac14 (\eta^{kb}\si^a_\m-\eta^{ka}\si^b_\m)
(\tau_\nu^\m \si^\nu_k- \tau^\nu\si^\m_{\nu k})
 L_{ab}{}^A{}_By^B -\tau^\nu y^A_\nu]\dr_A+
\nonumber\\
&& \qquad (\tau_\nu^\al k_\m{}^\nu{}_\bt -\tau_\bt^\nu
k_\m{}^\al{}_\nu -\tau_\mu^\nu k_\nu{}^\al{}_\bt +\tau_{\m\bt}^\al
-\tau^\nu k_{\nu\m}{}^\al{}_\bt)\frac{\dr}{\dr k_\mu{}^\al{}_\bt}.
\nonumber
\een
If infinitesimal gauge transformations are vector fields
(\ref{i16}), the configuration space of this gauge system is the
bundle product
\mar{i25}\beq
E=S\op\times_{\Si_T} C_K \op\times_{\Si_T}T_{\rL_s}P^0,
\label{i25}
\eeq
coordinated by $(x^\la,\si^\m_a, k_\la{}^\nu{}_\al,
y^A,\tau^\la,\xi^m)$, and the gauge symmetry is
\mar{i24}\ben
&& \up= (\tau_\nu^\al \si^\nu_c- \tau^\nu\si^\al_{\nu c} -\xi^m
I_m{}^d{}_c\si^\al_d) \dr^c_\al + (\xi^m I_m{}^A{}_B y^B-\tau^\nu
y^A_\nu)\dr_A +
\label{i24}\\
&& \qquad (\tau_\nu^\al k_\m{}^\nu{}_\bt -\tau_\bt^\nu
k_\m{}^\al{}_\nu -\tau_\mu^\nu k_\nu{}^\al{}_\bt +\tau_{\m\bt}^\al
-\tau^\nu k_{\nu\m}{}^\al{}_\bt)\frac{\dr}{\dr
k_\mu{}^\al{}_\bt}.\nonumber
\een

\section{BRST symmetries}

In order to introduce BRST symmetries, let us consider Lagrangian
systems of even and odd variables. We describe odd variables and
their jets on a smooth manifold $X$ as generating elements of the
structure ring of a graded manifold whose body is $X$
\cite{noether,cmp,ijmp}. This definition reproduces the heuristic
notion of jets of ghosts in the field-antifield BRST theory
\cite{barn,bran01}.

Recall that any graded manifold $(\gA,X)$ with a body $X$ is
isomorphic to the one whose structure sheaf $\gA_Q$ is formed by
germs of sections of the exterior product
\mar{g80}\beq
\w Q^*=\Bbb R\op\oplus_X Q^*\op\oplus_X\op\w^2
Q^*\op\oplus_X\cdots, \label{g80}
\eeq
where $Q^*$ is the dual of some real vector bundle $Q\to X$ of
fiber dimension $m$. In field models, a vector bundle $Q$ is
usually given from the beginning. Therefore, we consider graded
manifolds $(X,\gA_Q)$ where the above mentioned isomorphism holds,
and call $(X,\gA_Q)$ the simple graded manifold constructed from
$Q$. The structure ring $\cA_Q$ of sections of $\gA_Q$ consists of
sections of the exterior bundle (\ref{g80}) called graded
functions. Let $\{c^a\}$ be the fiber basis for $Q^*\to X$,
together with transition functions $c'^a=\rho^a_bc^b$. It is
called the local basis for the graded manifold $(X,\gA_Q)$. With
respect to this basis, graded functions read
\be
f=\op\sum_{k=0}^m \frac1{k!}f_{a_1\ldots a_k}c^{a_1}\cdots
c^{a_k},
\ee
where $f_{a_1\cdots a_k}$ are local smooth real functions on $X$.

Given a graded manifold $(X,\gA_Q)$, let $\gd\cA_Q$ be the
$\cA_Q$-module of $\Bbb Z_2$-graded derivations of the $\Bbb
Z_2$-graded ring of $\cA_Q$, i.e.,
\be
u(ff')=u(f)f'+(-1)^{[u][f]}fu (f'), u\in\gd\cA_Q, \qquad f,f'\in
\cA_Q,
\ee
where $[.]$ denotes the Grassmann parity. Its elements are called
$\Bbb Z_2$-graded (or, simply, graded) vector fields on
$(X,\gA_Q)$. Due to the canonical splitting $VQ= Q\times Q$, the
vertical tangent bundle $VQ\to Q$ of $Q\to X$ can be provided with
the fiber bases $\{\dr_a\}$ which is the dual of $\{c^a\}$. Then a
graded vector field takes the local form $u= u^\la\dr_\la +
u^a\dr_a$, where $u^\la, u^a$ are local graded functions. It acts
on $\cA_Q$ by the rule
\mar{cmp50'}\beq
u(f_{a\ldots b}c^a\cdots c^b)=u^\la\dr_\la(f_{a\ldots b})c^a\cdots
c^b +u^d f_{a\ldots b}\dr_d\rfloor (c^a\cdots c^b). \label{cmp50'}
\eeq
This rule implies the corresponding transformation law
\be
u'^\la =u^\la, \qquad u'^a=\rho^a_ju^j +
u^\la\dr_\la(\rho^a_j)c^j.
\ee
Then one can show that graded vector fields on a simple graded
manifold can be represented by sections of the vector bundle
$\cV_Q\to X$, locally isomorphic to $\w Q^*\ot_X(Q\oplus_X TX)$.

Accordingly, graded exterior forms on the graded manifold
$(X,\gA_Q)$ are introduced as sections of the exterior bundle
$\op\w\cV^*_Q$, where $\cV^*_Q\to X$ is the $\w Q^*$-dual of
$\cV_Q$. Relative to the dual local bases $\{dx^\la\}$ for $T^*X$
and $\{dc^b\}$ for $Q^*$, graded one-forms read
\be
\f=\f_\la dx^\la + \f_adc^a,\qquad \f'_a=\rho^{-1}{}_a^b\f_b,
\qquad \f'_\la=\f_\la +\rho^{-1}{}_a^b\dr_\la(\rho^a_j)\f_bc^j.
\ee
The duality morphism is given by the interior product
\be
u\rfloor \f=u^\la\f_\la + (-1)^{[\f_a]}u^a\f_a.
\ee
Graded exterior forms constitute the bigraded differential algebra
(henceforth BGDA) $\cC^*_Q$ with respect to the bigraded exterior
product $\w$ and the exterior differential $d$.

Since the jet bundle $J^rQ\to X$ of a vector bundle $Q\to X$ is a
vector bundle, let us consider the simple graded manifold
$(X,\gA_{J^rQ})$ constructed from $J^rQ\to X$. Its local basis is
$\{x^\la,c^a_\La\}$, $0\leq |\La|\leq r$, together with the
transition functions
\mar{+471}\beq
c'^a_{\la +\La}=d_\la(\rho^a_j c^j_\La), \qquad d_\la=\dr_\la +
\op\sum_{|\La|<r}c^a_{\la+\La} \dr_a^\La, \label{+471}
\eeq
where $\dr_a^\La$ are the duals of $c^a_\La$. Let $\cC^*_{J^rQ}$
be the BGDA of graded exterior forms on the graded manifold
$(X,\gA_{J^rQ})$. A linear bundle morphism $\pi^r_{r-1}:J^rQ \to
J^{r-1}Q$ yields the corresponding monomorphism of BGDAs
$\cC^*_{J^{r-1}Q}\to \cC^*_{J^rQ}$. Hence, there is the direct
system of BGDAs
\mar{g205}\beq
\cC^*_Q\ar^{\pi^{1*}_0} \cC^*_{J^1Q}\cdots
\ar^{\pi^r_{r-1}{}^*}\cC^*_{J^rQ}\ar\cdots. \label{g205}
\eeq
Its direct limit $\cC^*_\infty Q$ consists of graded exterior
forms on graded manifolds $(X,\gA_{J^rQ})$, $r\in\Bbb N$, modulo
the pull-back identification, and it inherits the BGDA operations
intertwined by the monomorphisms $\pi^r_{r-1}{}^*$. It is a
$C^\infty(X)$-algebra locally generated by the elements $(1,
c^a_\La, dx^\la,\th^a_\La=dc^a_\La -c^a_{\la +\La}dx^\la)$,
$0\leq|\La|$.

 In order to regard even and odd dynamic variables on the
same footing,  let $Y\to X$ be hereafter an affine bundle (or a
subbundle of an affine bundle which need not be affine), and let
$\cP^*_\infty Y\subset \cO^*_\infty Y$ be the
$C^\infty(X)$-subalgebra of exterior forms whose coefficients are
polynomial in the fiber coordinates $y^i_\La$ on jet bundles $J^r
Y\to X$. Let us consider the product
\mar{0670}\beq
\cS^*_\infty=\cC_\infty^*Q\w\cP^*_\infty Y \label{0670}
\eeq
of graded algebras $\cC_\infty^*Q$ and $\cP^*_\infty Y$ over their
common graded subalgebra $\cO^*X$ of exterior forms on $X$
\cite{cmp}. It consists of the elements
\be
\op\sum_i \psi_i\ot\f_i, \qquad \op\sum_i \f_i\ot\psi_i, \qquad
\psi\in \cC^*_\infty Q, \qquad \f\in \cP^*_\infty Y,
\ee
modulo the commutation relations
\mar{0442}\ben
&&\psi\ot\f=(-1)^{|\psi||\f|}\f\ot\psi, \qquad
\psi\in \cC^*_\infty Q, \qquad \f\in \cP^*_\infty Y, \label{0442}\\
&& (\psi\w\si)\ot\f=\psi\ot(\si\w\f), \qquad \si\in \cO^*X.
\nonumber
\een
They are  endowed with the total form degree $|\psi|+|\f|$ and the
total Grassmann parity $[\psi]$. Their multiplication
\mar{0440}\beq
(\psi\ot\f)\w(\psi'\ot\f'):=(-1)^{|\psi'||\f|}(\psi\w\psi')\ot
(\f\w\f'). \label{0440}
\eeq
obeys the relation
\be
\vf\w\vf' =(-1)^{|\vf||\vf'| +[\vf][\vf']}\vf'\w \vf, \qquad
\vf,\vf'\in \cS^*_\infty,
\ee
and makes $\cS^*_\infty$ (\ref{0670}) into a bigraded $C^\infty
(X)$-algebra.  For instance, elements of the ring $S^0_\infty$ are
polynomials of $c^a_\La$ and $y^i_\La$ with coefficients in
$C^\infty(X)$.

The algebra $\cS^*_\infty$ is provided with the exterior
differential
\mar{0441}\beq
d(\psi\ot\f):=(d_\cC\psi)\ot\f +(-1)^{|\psi|}\psi\ot(d_\cP\f),
\qquad \psi\in \cC^*_\infty, \qquad \f\in \cP^*_\infty,
\label{0441}
\eeq
where $d_\cC$ and $d_\cP$ are exterior differentials on the
differential algebras $\cC^*_\infty Q$ and $\cP^*_\infty Y$,
respectively. It obeys the relations
\be
 d(\vf\w\vf')= d\vf\w\vf' +(-1)^{|\vf|}\vf\w d\vf', \qquad
\vf,\vf'\in \cS^*_\infty,
\ee
and makes $\cS^*_\infty$ into a BGDA. We agree to call elements of
$\cS^*_\infty$ the graded exterior forms on $X$. Hereafter, let
the collective symbols $s^A_\La$ stand both for even and odd
generating elements $c^a_\La$, $y^i_\La$ of the $C^\infty(X)$-ring
$\cS^0_\infty$. Then the BGDA $\cS^*_\infty$ is locally generated
by $(1,s^A_\La, dx^\la, \th^A_\La=ds^A_\La -s^A_{\la+\La}dx^\la)$,
$|\La|\geq 0$. Since the generating elements $s^A_\La$ and
$\th^A_\La$ are derived from $s^A$, the set $\{s^A\}$ is called
the local basis for the BGDA $S^*_\infty$.

Similarly to $\cO^*_\infty Y$, the BGDA $\cS^*_\infty$ is
decomposed into $\cS^0_\infty$-modules $\cS^{k,r}_\infty$ of
$k$-contact and $r$-horizontal graded forms together with the
corresponding projections $h_k$ and $h^r$. Accordingly, the
exterior differential $d$ (\ref{0441}) on $\cS^*_\infty$ is split
into the sum $d=d_H+d_V$ of the total and vertical differentials
\be
d_H(\f)=dx^\la\w d_\la(\f), \qquad d_V(\f)=\th^A_\La\w\dr^\La_A
\f, \qquad \f\in \cS^*_\infty.
\ee
One can think of the elements
\be
L=\cL\om\in \cS^{0,n}_\infty, \qquad \dl (L)= \op\sum_{|\La|\geq
0}
 (-1)^{|\La|}\th^A\w d_\La (\dr^\La_A L)\in \cS^{0,n}_\infty
\ee
as being a graded Lagrangian and its Euler--Lagrange operator,
respectively.

A graded derivation  $\vt\in\gd \cS^0_\infty$ of the $\Bbb R$-ring
$\cS^0_\infty$ is said to be contact if the Lie derivative
$\bL_\vt$ preserves the ideal of contact graded forms of the BGDA
$\cS^*_\infty$. With respect to the local basis $(x^\la,s^A_\La,
dx^\la,\th^A_\La)$ for the BGDA $\cS^*_\infty$, any contact graded
derivation takes the form
\mar{g105}\beq
\vt=\vt_H+\vt_V=\vt^\la d_\la + (\vt^A\dr_A +\op\sum_{|\La|>0}
d_\La\vt^A\dr_A^\La), \label{g105}
\eeq
where $\vt^\la$, $\vt^A$ are local graded functions \cite{cmp}.
The interior product $\vt\rfloor\f$ and the Lie derivative
$\bL_\vt\f$, $\f\in\cS^*_\infty$, are defined by the formulae
\be
&& \vt\rfloor \f=\vt^\la\f_\la + (-1)^{[\f_A]}\vt^A\f_A, \qquad
\f\in \cS^1_\infty,\\
&& \vt\rfloor(\f\w\si)=(\vt\rfloor \f)\w\si
+(-1)^{|\f|+[\f][\vt]}\f\w(\vt\rfloor\si), \qquad \f,\si\in
\cS^*_\infty, \\
&& \bL_\vt\f=\vt\rfloor d\f+ d(\vt\rfloor\f), \qquad
\bL_\vt(\f\w\si)=\bL_\vt(\f)\w\si
+(-1)^{[\vt][\f]}\f\w\bL_\vt(\si).
\ee

The Lie derivative $\bL_\vt L$ of a Lagrangian $L$ along a contact
graded derivation $\vt$ (\ref{g105}) fulfills the first
variational formula
\mar{g107}\beq
\bL_\vt L= \vt_V\rfloor\dl L +d_H(h_0(\vt\rfloor \Xi_L)) + d_V
(\vt_H\rfloor\om)\cL, \label{g107}
\eeq
where $\Xi_L$ is a Lepagean equivalent of a graded Lagrangian $L$
\cite{cmp}.

A contact graded derivation $\vt$ is said to be variational if the
Lie derivative (\ref{g107}) is $d_H$-exact. A glance at the
expression (\ref{g107}) shows that: (i) a contact graded
derivation $\vt$ is variational only if it is projected onto $X$,
and (ii) $\vt$ is variational iff its vertical part $\vt_V$ is
well. Therefore, we restrict our consideration to vertical contact
graded derivations
\mar{0672}\beq
\vt=\op\sum_{0\leq|\La|} d_\La\up^A\dr_A^\La. \label{0672}
\eeq
Such a derivation is completely defined by its first summand
\mar{0673}\beq
\up=\up^A(x^\la,s^A_\La)\dr_A, \qquad 0\leq|\La|\leq k,
\label{0673}
\eeq
which is also a graded derivation of $\cS^0_\infty$. It is called
the generalized graded vector field. A glance at the first
variational formula (\ref{g107}) shows that $\vt$ (\ref{0672}) is
variational iff $\up\rfloor \dl L$ is $d_H$-exact.

A vertical contact graded derivation $\vt$ (\ref{0672}) is said to
be nilpotent if
\mar{g133}\beq
\bL_\up(\bL_\up\f)= \op\sum_{|\Si|\geq 0,|\La|\geq 0 }
(\up^B_\Si\dr^\Si_B(\up^A_\La)\dr^\La_A +
(-1)^{[s^B][\up^A]}\up^B_\Si\up^A_\La\dr^\Si_B \dr^\La_A)\f=0
\label{g133}
\eeq
for any horizontal graded form $\f\in S^{0,*}_\infty$ or,
equivalently, $(\vt\circ\vt)(f)=0$ for any graded function $f\in
\cS^0_\infty$. One can show that $\vt$ is nilpotent only if it is
odd and iff the equality
\mar{0688}\beq
\vt(\up^A)=\op\sum_{|\Si|\geq 0} \up^B_\Si\dr^\Si_B(\up^A)=0
\label{0688}
\eeq
holds for all $\up^A$ \cite{cmp}.

Return now to the original gauge system on a fiber bundle $Y$ with
a Lagrangian $L$ (\ref{0512}) and a gauge symmetry $\up$
(\ref{0509}). Let us consider the BGDA
\be
\cS^*_\infty[W;Y]=\cC^*_\infty W\w\cP^*_\infty Y
\ee
whose local basis is $\{c^r, y^i\}$. Let $L\in \cO^{0,n}_\infty Y$
be a polynomial in $y^i_\La$, $0\leq |L|$. Then it is a graded
Lagrangian $L\in \cP^{0,n}_\infty Y\subset \cS^{0,n}_\infty[W;Y]$
in $\cS^*_\infty[W;Y]$. Its gauge symmetry $\up$ (\ref{0509})
gives rise to the generalized vector field $\up_E=\up$ on $E$, and
the latter defines the generalized graded vector field $\up$
(\ref{0673}) by the formula (\ref{0680}). It is easily justified
that the contact graded derivation $\vt$ (\ref{0672}) generated by
$\up$ (\ref{0680}) is variational for $L$. It is odd, but need not
be nilpotent. However, one can try to find a nilpotent contact
graded derivation (\ref{0672}) generated by some generalized
graded vector field (\ref{0684}) which coincides with $\vt$ on
$\cP^*_\infty Y$. Then $\up$ (\ref{0684}) is  called a BRST
symmetry.

In this case, the nilpotency conditions (\ref{0688}) read
\mar{0690,1}\ben
&& \op\sum_\Si d_\Si(\op\sum_\Xi\up^{i,\Xi}_rc^r_\Xi)
\op\sum_\La\dr^\Si_i (\up^{j,\La}_s)c^s_\La +\op\sum_\La d_\La
(u^r)\up^{j,\La}_r
=0, \label{0690}\\
&& \op\sum_\La(\op\sum_\Xi d_\La(\up^{i,\Xi}_r c^r_\Xi)\dr^\La_i
+d_\La (u^r)\dr_r^\La)u^q=0\label{0691}
\een
for all indices $j$ and $q$. They are equations for graded
functions $u^r\in\cS^0_\infty[W;Y]$. Since these functions are
polynomials
\mar{0693}\beq
u^r=u_{(0)}^r + \op\sum_\G u_{(1)p}^{r,\G} c^p_\G +
\op\sum_{\G_1,\G_2} u_{(2)p_1p_2}^{r,\G_1\G_2}
c^{p_1}_{\G_1}c^{p_2}_{\G_2} +\cdots \label{0693}
\eeq
in $c^s_\La$, the equations (\ref{0690}) -- (\ref{0691}) take the
form
\mar{0694}\ben
&& \op\sum_\Si d_\Si(\op\sum_\Xi\up^{i,\Xi}_rc^r_\Xi)
\op\sum_\La\dr^\Si_i (\up^{j,\La}_s)c^s_\La +\op\sum_\La d_\La
(u_{(2)}^r)\up^{j,\La}_r
=0, \label{0694a}\\
&& \op\sum_\La d_\La (u_{(k\neq 2)}^r)\up^{j,\La}_r =0,
\label{0694b}\\
&& \op\sum_\La\op\sum_\Xi d_\La(\up^{i,\Xi}_r c^r_\Xi)\dr^\La_i
u_{(k-1)}^q +\op\sum_{m+n-1=k}d_\La (u_{(m)}^r)\dr_r^\La u_{(n)}^q
=0. \label{0694c}
\een
One can think of the equalities (\ref{0694a}) and (\ref{0694c})
 as being the generalized commutation relations and
generalized Jacobi identities of original gauge transformations,
respectively \cite{algebr}.

\section{BRST symmetries in gauge gravitation theories}

Following the procedure in Section 5, let us obtain BRST
symmetries of gauge gravitation theories listed in Section 2.

Given a gauge system on the fiber bundle (\ref{i20}), let us
consider the BGDA
\mar{i40}\beq
\cS^*_\infty[TX;\Si\op\times_X C_K]=\cC^*_\infty[TX]\w
\cP_\infty[\Si\op\times_X C_K], \label{i40}
\eeq
whose basis consists of even elements $\si^{\al\bt}$,
$k_\m{}^\al{}_\bt$ and odd elements $c^\la$, which are ghosts
substituting for parameters $\tau^\la$ in the gauge symmetry
(\ref{i21}). We obtain the nilpotent BRST symmetry
\mar{i30}\ben
&&\up=\up^{\al\bt}\frac{\dr}{\dr\si^{\al\bt}} +\up_\m{}^\al{}_\bt
\frac{\dr}{\dr k_\mu{}^\al{}_\bt} +\up^\la \frac{\dr}{\dr
c^\la}=(\si^{\nu\bt} c_\nu^\al +\si^{\al\nu}
c_\nu^\bt-c^\la\si_\la^{\al\bt})\frac{\dr}{\dr \si^{\al\bt}}+
\label{i30}\\
&& \qquad (c_\nu^\al k_\m{}^\nu{}_\bt -c_\bt^\nu k_\m{}^\al{}_\nu
-c_\mu^\nu k_\nu{}^\al{}_\bt +c_{\m\bt}^\al-c^\la
k_{\la\mu}{}^\al{}_\bt)\frac{\dr}{\dr k_\mu{}^\al{}_\bt} +
c^\la_\m c^\m\frac{\dr}{\dr c^\la}. \nonumber
\een

The BGDA of the gauge system on the fiber bundle (\ref{gr10}) is
$\cS^*_\infty[T_GP;C\times\Si]$ whose odd basis elements $c^r$ and
$c^\la$ substitute for parameters $\xi^r$ and $\tau^\la$ in the
gauge symmetry (\ref{gr11}). The corresponding BRST symmetry reads
\mar{i31}\ben
&& \up= (c^r_{pq}a^p_\la c^q + c^r_\la -a^r_\m c^\m_\la-c^\m
a_{\m\la}^r)\frac{\dr}{\dr a_\la^r}  + (\si^{\nu\bt} c_\nu^\al
+\si^{\al\nu} c_\nu^\bt-c^\la\si_\la^{\al\bt})\frac{\dr}{\dr
\si^{\al\bt}}
+ \label{i31}\\
&& \qquad (-\frac12c^r_{pq}c^pc^q -c^\m c^r_\m)\frac{\dr}{\dr c^r}
+ c^\la_\m c^\m\frac{\dr}{\dr c^\la}. \nonumber
\een

Turn now to the gauge theory of gravity in the presence of fermion
field. Since gauge transformations (\ref{i12}) fail to form a
closed algebra, one can not associate to them a nilpotent BRST
symmetry. Therefore, let us consider the gauge gravitation system
on the fiber bundle (\ref{i25}). Its BGDA is
\be
\cS^*_\infty[T_{\rL_s}P^0;S\op\times_{\Si_T} C_K],
\ee
whose odd basis elements $c^m$ and $c^\la$ play the role of ghosts
of the gauge symmetry (\ref{i24}). The associated BRST symmetry is
\mar{i32}\ben
&& \up=(c_\nu^\al \si^\nu_c- c^\nu\si^\al_{\nu c} -c^m
I_m{}^d{}_c\si^\al_d)\frac{\dr}{\dr\si_c^\al} + (c^m I_m{}^A{}_B
y^B-c^\nu y^A_\nu)\dr_A +
\label{i32}\\
&& \qquad (c_\nu^\al k_\m{}^\nu{}_\bt -c_\bt^\nu k_\m{}^\al{}_\nu
-c_\mu^\nu k_\nu{}^\al{}_\bt +c_{\m\bt}^\al-c^\la
k_{\la\mu}{}^\al{}_\bt)\frac{\dr}{\dr k_\mu{}^\al{}_\bt}
+\nonumber\\
&&\qquad (-\frac12c^m_{pq}c^pc^q -c^\m c^m_\m)\frac{\dr}{\dr c^m}
+ c^\la_\m c^\m\frac{\dr}{\dr c^\la}. \nonumber
\een

\section{The BV quantization}

As was mentioned above, we restrict our consideration to the BV
quantization of the metric-affine gravitation theory. Its original
Lagrangian $L_{\rm MA}$ on the fiber bundle $\Si\times C_K$ need
not be specified, but it possesses the gauge symmetry $\up$
(\ref{i21}). It is important that this gauge symmetry is complete
and irreducible.

Let us note that any Lagrangian $L$ has gauge symmetries. In
particular, there always exist trivial gauge symmetries
\be
\up=\op\sum_\La \eta(M)^{i,\La}_r\xi^r_\La, \qquad
M^{i,\La}_r=\op\sum_\Si T^{i,j,\La,\Si}d_\Si\cE_j,  \qquad
T_r^{j,i,\La,\Si}=-T_r^{i,j,\Si,\La}.
\ee
Furthermore, given a gauge symmetry $\up$ (\ref{0509}), let $h$ be
a linear differential operator on some vector bundle $E'\to Y$,
coordinated by  $(x^\la,y^i,\xi'^s)$, with values in the vector
bundle $E$. Then the composition $\up\circ
h=\up'^{i,\La}_s\xi'^s_\La\dr_i$ is a variational symmetry of the
pull-back onto $E'$ of a Lagrangian $L$ on $Y$, i.e., a gauge
symmetry of $L$. In view of this ambiguity, we agree to say that a
gauge symmetry $\up$ (\ref{0509}) of a Lagrangian $L$ is complete
if a different gauge symmetry $\up'$ of $L$ factors through $\up$
as
\be
\up'_0=\up\circ h + T, \qquad T\ap 0.
\ee

In the case of a generic metric-affine Lagrangian $L_{\rm MA}$ on
the jet manifold $J^1(\Si\times C_K)$, the gauge symmetry $\up$
(\ref{i21}) is complete. Although there are Lagrangians, e.g., the
Hilbert--Einstein scalar curvature and the Yang--Mills type
Lagrangian which admit more wide gauge symmetries. Note that a
metric-affine Lagrangian $L_{\rm MA}$ usually depends on
$k_\m{}^\al{}_\bt$ through the curvature $R_{\la\m}{}^\al{}_\bt$
(\ref{i52a}). The Euler--Lagrange operator
\mar{i38}\beq
\dl L_{\rm MA}=\cE_{\al\bt} d\si^{\al\bt}\w\om +
\cE^\m{}_\al{}^\bt dk_\m{}^\al{}_\bt\w\om \label{i38}
\eeq
of a metric-affine Lagrangian takes its values in the fiber bundle
\mar{i36}\beq
(V^*\Si\op\times_X V^*C_K)\op\ot_{\Si\times C_K}\op\w^4T^*X,
\label{i36}
\eeq
coordinated by $(x^\la,\si^{\m\nu}, k_\la{}^\nu{}_\al,
\ol\si_{\m\nu},\ol k^\la{}_\nu{}^\al)$. Due to the canonical
splittings
\mar{i45}\beq
V^*\Si=\Si\op\times_X \Si^*=\Si\op\times_X \op\vee^2TX, \qquad
V^*C_K=C_K\op\times_X C_K^*=C_K\op\times_X( \op\ot^2TX\op\ot_X
T^*X), \label{i45}
\eeq
the fiber bundle (\ref{i36}) is isomorphic to
\be
(\Si\op\times_X C_K)\op\times_X(\Si^*\op\times_X C^*_K)\op\ot_X
\op\w^4T^*X.
\ee

A complete gauge symmetry $\up\not\ap 0$ (\ref{0509}) is called
reducible if  there exist a vector bundle $W_0\to X$ and a
$E$-valued differential operator $\up^0\not\ap 0$  on the bundle
product $E_0=Y\times W_0$ which is linear on $W_0$ and obeys the
equality $\up\circ \up^0\ap 0$. If $\up'^0$ is another
differential operator possessing these properties, then $\up'^0$
factors through $\up^0$.

The gauge symmetry (\ref{i21}) is irreducible because its term of
maximal jet order is $\tau^\al_{\m\bt}\dr_\al^{\m\bt}$. Indeed,
let us assume that there exist a vector bundle $W_0\to X$
coordinated by $(x^\la,w^A)$ and  the above mentioned differential
operator $\up^0$ on the bundle product
\be
E_0=\Si\op\times_X C_K\op\times_X W_0
\ee
with values in the vector bundle $E$ (\ref{i20}) such that
$\up\circ \up^0\ap 0$. Given the maximal jet order term
\mar{i33}\beq
\op\sum_{|\Xi|=k} \up_0{}^{\al,\Xi}_A w^A_\Xi\frac{\dr}{\dr
\tau^\al}\not\ap 0 \label{i33}
\eeq
of the operator $\up^0$, the equality $\up\circ \up^0\ap 0$
results in the condition
\be
\op\sum_{|\Xi|=k} \up_0{}^{\al,\Xi}_A
d_{\m\bt}w^A_\Xi\frac{\dr}{\dr k_\m{}^\al{}_\bt}\ap 0.
\ee
It follows that all coefficients $\up_0{}^{\al,\Xi}_A$ and,
consequently, the term (\ref{i33}) vanish on-shell.

By virtue of Noether's second theorem, the gauge symmetry
(\ref{i21}) defines the Noether operator $\Delta$ (\ref{0511}) on
the fiber bundle (\ref{i36}) with values in the fiber bundle
\mar{i37}\beq
(\Si\op\times_X C_K\op\times_X T^*X)\op\ot_{\Si\times
C_K}\op\w^4T^*X, \label{i37}
\eeq
coordinated by $(x^\la,\si^{\m\nu}, k_\la{}^\nu{}_\al,
\ol\tau_\la)$. The formula (\ref{0657}) gives
\mar{i35}\ben
&& \Delta=\tau^\la\op\sum_{|\La|=0,1,2}[
\Delta_\la^{\al\bt,\La}d_\La\ol\si_{\al\bt} +
\Delta_{\la\mu}{}^\al{}_\bt{}^{,\La}d_\La \ol k^\m{}_\al{}^\bt]\om
= \label{i35}\\
&& \quad \tau^\la[-(\si^{\al\bt}_\la
+2\si^{\nu\bt}_\nu\dl^\al_\la)\ol\si_{\al\bt}
-2\si^{\nu\bt}d_\nu\ol\si_{\la\bt} +(-k_{\la\m}{}^\al{}_\bt
-k_{\nu\m}{}^\nu{}_\bt\dl^\al_\la + k_{\bt\m}{}^\al{}_\la +
k_{\m\la}{}^\al{}_\bt)\ol k^\m{}_\al{}^\bt +\nonumber\\
&& \quad (-k_\m{}^\nu{}_\bt\dl^\al_\la
+k_\m{}^\al{}_\la\dl^\nu_\bt +k_\la{}^\al{}_\bt\dl^\nu_\m)d_\nu\ol
k^\m{}_\al{}^\bt + d_{\m\bt} \ol k^\m{}_\la{}^\bt]\om. \nonumber
\een

Following the standard BV quantization procedure \cite{bat,gom},
let us construct the extended Lagrangian which obey certain
condition and depends on ghosts and antifields.

In order to introduce antifields of even fields $\si^{\al\bt}$,
$k_\m{}^\al{}_\bt$ and odd ghosts $\tau^\la$, let us extend the
BGDA (\ref{i40}) to the BGDA
\mar{i41}\beq
\cS^*_\infty[TX\op\times_X (\Si^*\op\times_X
C^*_K)\op\ot_X\op\w^4T^*X ;\Si\op\times_X C_K\op\times_X
T^*X\op\ot_X\op\w^4T^*X], \label{i41}
\eeq
where the splittings (\ref{i45}) has been used. The local basis
for the BGDA (\ref{i41}) is
\mar{i46}\beq
\{\si^{\m\nu}, k_\m{}^\al{}_\bt, c^\la, \ol\si_{\m\nu},\ol
k^\m{}_\al{}^\bt, \ol c_\la\}, \label{i46}
\eeq
where the antifields $\ol\si_{\m\nu}$, $\ol k^\la{}_\nu{}^\al$ are
odd, while the antifields $\ol c_\la$ are even. It is convenient
to use right graded derivations $\rdr$ with respect to antifields.
They act on graded functions and forms $\f$ on the right by the
rule
\be
\rdr(\f)=d\f\lfloor \rdr +d(\f\lfloor \rdr), \qquad
\rdr(\f\w\f')=(-1)^{[\f']}\rdr(\f)\w\f'+ \f\w \rdr(\f').
\ee

An original metric-affine Lagrangian $L_{\rm MA}$ is assumed to be
polynomial in variables $\si^{\m\nu}$, $k_\la{}^\nu{}_\al$.
Therefore, it is obviously a graded Lagrangian in the BGDA
(\ref{i41}). Given the BRST symmetry $\up$ (\ref{i30}), let us
consider the graded density
\mar{i50}\beq
L=\cL\om= L_{\rm MA} + \up^{\al\bt}\ol\si_{\al\bt}\om +
\up_\m{}^\al{}_\bt \ol k^\m{}_\al{}^\bt\om + \up^\la \ol c_\la\om.
\label{i50}
\eeq
One can show that it obeys the following conditions
\mar{i51}\ben
&& \up^\al_c =\frac{\rdr\cL }{\dr \ol\si^c_\al}, \qquad
\up_\m{}^\al{}_\bt= \frac{\rdr\cL }{\dr \ol k^\m{}_\al{}^\bt},
\qquad \up^\la= \frac{\rdr\cL}{\dr \ol c_\la}, \label{i51a}\\
&& \bL_\up L= \frac{\rdr\cL }{\dr \ol\si^c_\al}\frac{\dl L}{\dl
\si^\al_c}+ \frac{\rdr\cL }{\dr \ol k^\m{}_\al{}^\bt} \frac{\dl
L}{\dl k_\m{}^\al{}_\bt} + \frac{\rdr\cL}{\dr \ol c_\la}
\frac{\dl L}{\dl c^\la} +d_H\si=d_H\si',\label{i51b}\\
&& L|_{\ol\si=\ol k=\ol c=0}=L_{\rm MA}. \label{i51c}
\een
A glance at these conditions shows that $L$ (\ref{i50}) is a
proper solution of the master equation (\ref{i51b})
\cite{bat,gom}. Thus, it is the desired extended Lagrangian.

One can write different variants of a gauged-fixed Lagrangian
(see, e.g., \cite{kaz} for the problem of the general covariance
in quantum gravity). By analogy with the Lorenz gauge in
Yang--Mills gauge theory, let us choose the gauge-fixing quantity
\mar{i53}\beq
\cS^\al_\bt=\si^{\la\m}S_{\la\m}{}^\al{}_\bt
=\si^{\la\m}(k_{\la\m}{}^\al{}_\bt+ k_{\m\la}{}^\al{}_\bt),
\label{i53}
\eeq
where $S_{\la\m}{}^\al{}_\bt$ is given by the expression
(\ref{i52b}). In order to define a gauge-fixing density, we add
even elements $B^\al_\bt$, odd elements $C_\al^\bt$ and their even
antifields $\ol C^\al_\bt$ to the local basis for the BGDA
(\ref{i41}). Then this gauge-fixing density reads
\mar{i54}\beq
\Phi= C_\al^\bt(\frac12 B^\al_\bt - \cS^\al_\bt)\om. \label{i54}
\eeq
The desired gauged-fixed Lagrangian $L_{\rm GF}$ comes from the
graded Lagrangian
\be
L+B^\al_\bt\ol C_\al^\bt\om= L_{\rm MA} +
\up^{\al\bt}\ol\si_{\al\bt}\om + \up_\m{}^\al{}_\bt \ol
k^\m{}_\al{}^\bt\om + \up^\la \ol c_\la\om +B^\al_\bt\ol
C_\al^\bt\om
\ee
by means of replacement of antifields with the variational
derivatives
\be
&&\ol\si_{\al\bt}=\frac{\dl\Phi}{\dl
\si^{\al\bt}}=-C^\m_\nu(k_{\al\bt}{}^\nu{}_\m+
k_{\bt\al}{}^\nu{}_\m) , \qquad \ol
k^\m{}_\al{}^\bt=\frac{\dl\Phi}{\dl
k_\m{}^\al{}_\bt}=2d_\la(C^\bt_\al\si^{\la\m}), \\
&& \ol c_\la = \frac{\dl\Phi}{\dl c^\la}=0, \qquad \ol C^\bt_\al=
\frac{\dl\Phi}{\dl C^\al_\bt}=(\frac12 B^\bt_\al - S^\bt_\al).
\ee
One obtains
\mar{i55}\ben
&& \cL_{\rm GF}= \cL_{\rm MA} -
\up^{\al\bt}C^\m_\nu(k_{\al\bt}{}^\nu{}_\m+ k_{\bt\al}{}^\nu{}_\m)
+ 2\up^\m{}_\al{}^\bt d_\g(C^\bt_\al\si^{\g\m})  +B^\al_\bt
(\frac12 B^\bt_\al - S^\bt_\al)=\label{i55}\\
&& \qquad \cL_{\rm MA} - (\si^{\la\bt} c_\la^\al +\si^{\al\la}
c_\la^\bt-c^\la\si_\la^{\al\bt})C^\m_\nu(k_{\al\bt}{}^\nu{}_\m+
k_{\bt\al}{}^\nu{}_\m) +\nonumber\\
&& \qquad 2(c_\nu^\al k_\m{}^\nu{}_\bt -c_\bt^\nu k_\m{}^\al{}_\nu
-c_\mu^\nu k_\nu{}^\al{}_\bt +c_{\m\bt}^\al-c^\la
k_{\la\mu}{}^\al{}_\bt) d_\g(C^\bt_\al\si^{\g\m})+ \nonumber\\
&& \qquad  B^\al_\bt (\frac12 B^\bt_\al -
\si^{\la\m}(k_{\la\m}{}^\bt{}_\al + k_{\m\la}{}^\bt{}_\al)).
\nonumber
\een

Given the Lagrangian (\ref{i55}), one can not automatically write
the generating functional of quantum gravity because the BV
quantization procedure fails to provide a functional measure.
However, this generating functional at least is a Gaussian
integral with respect to variables $B^\al_\bt$. Therefore, we can
replace the Lagrangian (\ref{i55}) with the effective one which
takes the familiar form
\be
\cL_{\rm EGF}= \cL_{\rm MA} +C^\al_\bt \cM^\bt_{\al\la}c^\la
+\frac12 \si^{\la\m}\si^{\g\nu}(k_{\la\m}{}^\bt{}_\al +
k_{\m\la}{}^\bt{}_\al)(k_{\g\nu}{}^\al{}_\bt +
k_{\nu\g}{}^\al{}_\bt),
\ee
where $\cM$ is a linear third order differential operator acting
on ghosts $c^\la$.

\end{document}